%Paper: cond-mat/9408016
%From: gabler@guinness.ias.edu (Margaret Gabler)
%Date: Thu, 4 Aug 94 10:03:45 EDT

\input phyzzx

\nonstopmode
\nopubblock
\sequentialequations
\twelvepoint
\overfullrule=0pt
\tolerance=5000
\input epsf

\line{\hfill }
\line{\hfill PUPT 1489, IASSNS 94/59}
\line{\hfill cond-mat/}
\line{\hfill July 1994}

\titlepage
\title{Renormalization Group Approach to Low Temperature Properties of a
Non-Fermi Liquid Metal}

\author{Chetan Nayak\foot{Research supported in part by a Fannie
and John Hertz Foundation fellowship.~~~
nayak@puhep1.princeton.edu}}
\vskip .2cm
\centerline{{\it Department of Physics }}
\centerline{{\it Joseph Henry Laboratories }}
\centerline{{\it Princeton University }}
\centerline{{\it Princeton, N.J. 08544 }}

\author{Frank Wilczek\foot{Research supported in part by DOE grant
DE-FG02-90ER40542.~~~WILCZEK@IASSNS.BITNET}}
\vskip.2cm
\centerline{{\it School of Natural Sciences}}
\centerline{{\it Institute for Advanced Study}}
\centerline{{\it Olden Lane}}
\centerline{{\it Princeton, N.J. 08540}}
\endpage

\REF\hlr{B. Halperin, P. Lee, and N. Read, Phys. Rev. {\bf B47}
(1993) 7312.}

\REF\expts{R. Willet, M. Paalanen, R. Ruel, K. West, L.
Pfeiffer, and R. Bishop, Phys. Revl Lett. {\bf 65} (1990) 112;
R. Du, H. Stormer, D. Tsui, L. Pfeiffer, and K. West, Phys. Rev.
Lett. {\bf 70} (1993) 2994; R. Du, H. Stormer, D. Tsu, L. Pfeiffer, and
K. West, Solid State Comm. {\bf 90} (1994) 71 and references therein;
D. Leadley, R. Nicholas, C. Foxon, and J. Harris, Phys. Rev. Lett.
{\bf 72} (1994) 1906; V. Goldman, B. Su, and J. Jain, Phys. Rev. Lett.
{\bf 72} (1994) 2065.}

\REF\aswz{D. Arovas, R. Schrieffer, F. Wilczek. and A. Zee,
Nuclear Physics {\bf B251} (1985) 117; see especially p. 123.}

\REF\laugh{R. Laughlin, Phys. Rev. Lett. {\bf 60} (1988) 2677;
Science {\bf 242} (1988) 525; A. Fetter, C. Hanna, and R. Laughlin,
Phys. Rev. {\bf B39} (1989) 9679.}

\REF\cwwh{Y. Chen, F. Wilczek, E. Witten, and B. Halperin,
Int. J. Mod. Phys. {\bf B3} (1989) 1001.}

\REF\stat{J. Leinaas and J. Myrheim, Nuovo Cimento
{\bf 37B} (1977) 1; G. Goldin, R. Menikoff, and D. Sharp,
J. Math. Phys. {\bf 22} (1981) 1664; F. Wilczek,
Phys. Rev. Lett. {\bf 48} (1982); {\it ibid}. {\bf 49} (1982 957.}

\REF\morestat{F. Wilczek, A. Zee, Phys. Rev. Lett. {\bf 51} (1982)
2250; see
also the discussion and
several of the reprints in F. Wilczek, ed. {\it Fractional
Statistics and Anyon Superconductivity} (World Scientific,
Singapore 1990); and in addition especially
S. Zhang, H. Hanson, and S. Kivelson Phys. Rev. Lett. {\bf
62} (1989) 82; {\it ibid}. 980.}

\REF\jain{J. Jain, Phys. Rev. Lett. {\bf 63} (1989) 199;
Phys. Rev. {\bf B40} (1989) 8079; {\bf B41} (1990) 7653.}

\REF\fwg{F. Wilczek, in {\it Fractional Statistics and Anyon
Superconductivty}, {\it ibid}. pp. 80-88; M. Greiter and F. Wilczek,
Mod. Phys. Lett. {\bf B4} (1990) 1063.}

\REF\abrahams{E. Abrahams, {\it Beyond the Fermi Liquid\/}
Rutgers preprint (unpublished, 1993).}

\REF\varma{C. Varma, {\it Theoretical Framework for the Normal
State of Copper-Oxide Metals\/} Bell preprint, to appear
Los Alamos Symposium on Strong Correlations, 1993, ed. K. Bedell
{\it et. al}. (Addison-Wesley 1994).}

\REF\tsui{R. Du, H. Stormer, D. Tsui, A. Yeh, L. Pfeiffer, and
K. West, {\it Drastic Enhancement of Composite Fermion
Mass Near Landau Level Filling $\nu = 1/2$}, MIT-Bell Labs-
Princeton preprint (unpublished, 1994). }

\REF\polchgauge {J. Polchinski, {\it
Low-Energy Dynamics of the Spinon-Gauge
System\/} ITP preprint NSF-ITP-93-33 (1993)}

\REF\fixedpoint {C. Nayak and F. Wilczek, Nucl. Phys. {\bf B417} (1994)
359.}

\REF\physprop {C. Nayak and F. Wilczek, {\it Physical Properties of
Metals from a Renormalization Group Standpoint\/},
PUPT 1488, IAS 94/60.}

\REF\millis {B.L. Altschuler, L.B. Ioffe, and A.J. Millis,
{\it On the Low Energy Properties of Fermions with Singular
Interactions\/} MIT-Rutgers-Bell Labs preprint (unpublished, 1994).}

\REF\kim {Y.B. Kim, Furasaki, X.-G.Wen, and P.A. Lee,
{\it Gauge-Invariant Response Functions of Fermions Coupled
to a Gauge Field\/} MIT preprint (unpublished, 1994).}

\REF\altschuler{L.B. Ioffe, D.Lidsky, and B.L. Altschuler
Rutgers-MIT preprint (unpublished, 1994).}

\REF\kwon {H.J Kwon, A. Houghton,and J.B. Marston
{\it Gauge Interactions and Bosonized Fermi Liquids\/}
Brown preprint (unpublished, 1994).}

\REF\varma {C.Varma {\it Theory of the Copper-Oxide Metals\/}
Bell reprint (unpublished, 1994).}

\REF\kveschenko {D.V. Khveschenko and P.C.E. Stamp,
Phys. Rev. Lett. {\bf 71}, (1993) 2118.}

\REF\shankar {R. Shankar, Physica {\bf A177} (1991) 530; Rev. Mod.
Phys. {\bf 68} (1994) 129.}

\REF\polchint{J. Polchinski,
``Effective Field Theory and the Fermi Surface,''
Proceedings of the
1992 Theoretical Advanced Study Institute in Elementary Particle
Physics, ed. J. Harvey and J. Polchinski (World Scientific, Singapore,
1993).}

\REF\ben {G. Benfatto and G. Gallavotti, J. Stat. Phys. {\bf 59} (1990) 541;
Phys. Rev. {\bf B42} (1990) 9967.}

\REF\ma{As reviewed in S. Ma, {\it Modern
Theory of Critical Phenomena},
(Benjamin, Reading Mass. 1976).}

\REF\haldanelutt{F. D. M. Haldane, J. Phys. {\bf C14}
(1981) 2585.}

\REF\wenlutt{Reviewed in X.-G. Wen, Int. J. Mod. Phys. {\bf B6}
(1992) 1711.}

\REF\mahan{Reviewed in G. Mahan, {\it Many Body Physics}
(Plenum, New York 1981).  For a modern treatment, with new
insights, see A. Houghton, H. Kwon, J. Marston, and
R. Shankar, {\it Coulomb Interaction and the Fermi Liquid State: Solution
by Bosonization\/}  Brown preprint (1994) (to appear in
{\it J. Phys. C: Condensed Matter}).}

\REF\he {S. He, P.M. Platzman, and B.I. Halperin, Phys. Rev. Lett.
{\bf 69} (1992) 3804.}

\REF\instantonspecfun {Y. B. Kim and X.-G. Wen, {\it
Instantons and the spectral function of electrons in the
half-filled Landau level} MIT preprint 1994}

\REF\anderson{G. Baskaran and P. Anderson, Phys. Rev.
{\bf B37} (1988) 580; P. Anderson, Phys. Rev. Lett.
{\bf 64} (1990) 1839.}

\REF\gaugeguys{See especially
P. Lee and N. Nagaosa, Phys. Rev. {\bf B46}
(1992) 5621.}

\abstract{We expand upon
on an earlier renormalization group
analysis of a non-Fermi liquid fixed point that
plausibly govers
the two dimensional electron liquid in a magnetic
field near filling fraction $\nu=1/2$.
We give a more complete description of our somewhat
unorthodox
renormalization group transformation by
relating both our field-theoretic approach to a direct
mode elimination and our anisotropic scaling to the
general problem of incorporating curvature of the
Fermi surface.  We derive physical
consequences of the fixed point by showing how
they follow from renormalization group equations
for finite-size scaling, where the size may be set
by the temperature or by the frequency of interest.
In order fully to exploit this approach, it is necessary
to take into account composite operators, including in
some cases dangerous ``irrelevant'' operators.  We devote
special attention to gauge invariance, both
as a formal requirement and in its positive role
providing Ward identities constraining the
renormalization of composite operators.  We emphasize
that new considerations arise in describing properties of
the physical electrons (as opposed to the quasiparticles.)
We propose an experiment which, if feasible, will allow the
most characteristic feature of our results, that is
the divergence of the effective mass of the
quasiparticle near the nominal Fermi surface,
to be tested directly.
Some comparison with other recent, related
work is attempted.}

\endpage

\chapter{Introduction}

%refdump
In remarkable work,
Halperin, Lee, and Read [\hlr ] have developed a theory of
the two-dimensional electron gas that has gained some important
experimental support [\expts ].  Their theory is
based on the idea, suggested in
the early literature of anyon physics [\aswz ] and used to great
effect in the theory of anyon superconductivity [\laugh ,\cwwh ],
of approximating
the effect of quantum
statistics in an assembly of identical particles by a
uniform magnetic field.  Recall that in 2+1 dimensions one can transmute
the statistics of particles [\stat ] by attaching fictitious charge and
flux to them, or equivalently by coupling them to a Chern-Simons gauge
field [\morestat ].   The long-range part of the fictitious
gauge field (that is, the vector potential) accruing to
an assembly of many identical particles simply tracks the number of
particles inside, according to Stokes' theorem, because each particle
contributes a definite amount of flux.  Thus one may remove the
longest-range part of the statistical vector potential
by replacing it with that of a uniform magnetic field, hoping to
treat the residual part as a regular, essentially {\it local\/} and
therefore non-singular, perturbation.
Precisely at $\nu=1/2$ the background fictitious field thus introduced
cancels the real external magnetic field, suggesting that at this filling
factor the electrons can be treated as {\it free\/} fermions coupled
to the residual gauge field.
Jain [\jain ] has also fruitfully
emphasized, from a somewhat different point of view, the importance of
representing electrons as particle-fictitious flux composites,
and the special
significance of filling factors where the
real and fictitious flux cancel.  A general view of the
phase diagram in the magnetic field-statistics plane incorporating
these insights, consistently
founded on the
idea that generic
{\it small, local\/} perturbations on systems with a gap
(or perhaps even systems
having small phase space for low-energy excitations, as
around a Fermi surface) do not change their qualitative properties,
has been proposed [\fwg ].

Besides its phenomenological success in a somewhat esoteric
corner of condensed matter physics, and its profitable use
of
glamorous theoretical ideas, there is another important reason
to be interested in the theory of Halperin, Lee, and Read:
it gives us the first clearly formulated
example of a {\it non-Fermi liquid metal\/}  outside of 1+1 dimensions.
There is significant evidence that the copper-oxide superconductors
are, in their normal state, 2+1-dimensional non-Fermi liquid metals,
as Anderson has advocated forcefully  for several years now;
for recent reviews see
[\abrahams ] and [\varma ].
For all these reasons, it seems important to examine the theory closely,
and to develop techniques for treating it more rigorously.

The
original calcuations of Halperin, Lee, and Read were essentially
sophisticated perturbative calculations. (In this context
by sophisticated perturbation theory we mean, for
instance, that appropriate self-energies, rather than, say,
propagators are
calculated perturbatively.  In Feynman graphs, this amounts to summation
of selected infinite sums of graphs, \eg\ rainbows.)  However the
relevant coupling constant is not small,
and it is unclear {\it a priori\/}
why the calculations work as well as they do.  It comes, perhaps,
more as a relief than a surprise that some recent measurements do
not seem to agree with the perturbative results, even qualitatively
[\tsui ].
These are measurements of the effective mass as a function of deviation
from half-filling, a quantity which (we shall argue) is plausibly
sensitive to the running of the gauge coupling.  The running of the
gauge coupling is
an effect that is not included
in the original calculations.

Several approaches to improving the original calculations have
been proposed in the literature [\polchgauge - \kveschenko ].
We shall discuss them further, and
especially their relation to the approach adopted here and in
our previous work [\fixedpoint], in our concluding remarks.

Our work is based
on applying conventional renormalization group ideas to the coupled
fermion-Chern-Simons system including, importantly, an intrinsic
long-range fermion-fermion interaction.  We find an
infrared fixed point that plausibly governs the infrared behavior
for the Hall effect
near $\nu=1/2$.  This analysis forms a direct extension of
a similar
approach to Fermi liquid theory that has been
extensively developed
recently
[\shankar,\polchint,\ben]. Indeed, our fixed point
merges into the Fermi liquid fixed point, which
is simply effective gauge coupling
$\rightarrow 0$, when the intrinsic Fermion repulsion is
sufficiently long-range.  (Of course, if the interactions
are sufficiently singular they could in themselves
spoil conventional Fermi liquid behavior.)
In the interesting critical case
of $1/|k|$ interactions (as one has due to real -- \ie\
electromagnetic -- Coulomb repulsion),
the approach to zero coupling
is logarithmic.

In this paper, we revisit our non-Fermi liquid fixed
point.  Our goals in doing this are three-fold.  First, we
want to show that the somewhat unconventional aspects of our
earlier formulation, specifically the use of singularities in
dimensional regularization to identify renormalizations, and
the use of anisotropic scaling, are not essential -- everything
can be done in a conventional mode-elimination scheme.  We shall
also discuss the technical issue of
gauge invariance in a bit more detail.
Second, we will calculate more -- specifically, the
anomalous dimensions
of operators corresponding to
marginal perturbations of Fermi liquid theory:
the Landau parameters, impurity scattering, and the Cooper
instability channel.
We show that the Fermi liquid parameters
remain marginal
as a result of the Ward identities,
despite 1-loop corrections which are unique to the
non-Fermi liquid.  Impurity scattering and the Cooper
instability exhibit more interesting behaviors.
Third, we want to use the developed machinery to derive
physical consequences.
The most fundamental of these, that tests the most characteristic
property directly, concerns the speed of ballistic propagation
of quasiparticles.
We compute in addition the
temperature dependence of various thermodynamic properties
and transport coefficients from simple renormalization
group equations and finite-size scaling. These methods
are explicated in the context of Fermi and Luttinger
liquid theory in a companion paper [\physprop].
In this connection we will
emphasize an important point that we treated very sloppily
in [\fixedpoint ], namely that the quasiparticles are
fundamentally different
objects from the electrons, a fact that drastically
affects the calculation
of some physical quantities while making hardly any
difference
for others.

\chapter {A Renormalization Group Manifesto}

In [\fixedpoint], a somewhat unusual scaling was used
in the renormalization group procedure: momenta
perpendicular and parallel to the Fermi surface were
scaled differently. In order to elucidate the logic behind
this scaling, we will consider the case of a flat Fermi
surface, relevant to the ${k_F}\rightarrow\infty$ limit
and perhaps to the Fermi surfaces produced by tight-binding
Hamiltonians. In this context, we explore the freedom available
in the definition of the renormalization group. We then
restore curvature to the Fermi surface and show that
the scaling of [\fixedpoint] is the correct one for this problem.
The invariance of the effective Lagrangian under this scaling,
together with the one-loop calculation of renormalization
functions, may be used to write the scaling form of the
free energy. The temperature dependence is determined
by finite-size scaling, where the inverse temperature,
$\beta$, is the ``size'' in the time direction.

Since we will be using renormalization group transformations
of a different flavor from those familiar in other
contexts,
it is useful to review the
basic requirements that such a transformation must satisfy:

1. High momentum degrees of freedom should be removed.
Their effect is retained only in their contribution
to the effective Lagrangian for the low-energy degrees of
freedom. This
step (and all others) must be non-singular. In a perturbative
scheme, this is typically done by evaluating graphs
with the momenta in some directions
on internal lines
restricted to a narrow range at the cutoff. No external
legs on these graphs, and hence no fields in the low-energy
effective Lagrangian, may have momenta in this range.
There is considerable freedom in this choice. For instance,
one can eliminate a shell
$\Lambda-d\Lambda<{({k_x^2}+{k_y^2})^{1/2}}<\Lambda$
or simply $\Lambda-d\Lambda<{k_x}<\Lambda$. In either case,
the denominators of internal propagators cannot become
too small, so the procedure is non-singular.

2. The momenta should be rescaled so that the cutoff(s)
are returned to their original values.
In general, some of the momentum directions will be unrestricted
in internal loops and in the low-energy Lagrangian
(as the ${k_y}$ integration is in the
second mode elimination scheme above).
If the loop integrals are insensitive to the cutoffs in
these directions, it is possible to simply
take these cutoffs to infinity. These directions
may then be freely rescaled. For instance, in the theory
of dynamic critical phenomena [\ma ] one integrates out
high $k$ modes but not high $\omega$ modes. Internal loops
must have $\Lambda-d\Lambda<k<\Lambda$ and external legs
must have $k<\Lambda-d\Lambda$ while both have $-\infty<\omega<\infty$.
However, both $k$ and $\omega$ are scaled, $k\rightarrow sk$,
$\omega\rightarrow {s^z}\omega$.

3. The fields should be rescaled so that the quadratic part
of the effective Lagrangian is returned to its
original form. In general, it may not be possible
to return all of the quadratic terms to their original form.
In any given kinematical regime, some of the quadratic terms will
set the scale for the important fluctuations; these
are the terms which should be returned to their
original form. The other terms will either grow or scale
to zero as the renormalization group transformation is iterated. If they
scale to zero, they may be ignored at low energy. If they
grow, then they eventually become the important terms
which set the scale for fluctuations and the field rescaling
should be modified to preserve them.

Although strictly speaking it is not part of the definition of
the renormalization group, it is nevertheless
important to keep track of the symmetries of the
problem. If one is looking for a fixed point which exhibits a certain
symmetry, then one should choose a scaling in step 2
which respects this symmetry. It is also important to realize that
there is not a one-to-one correspondence between the
ways of carrying out steps 1 and 2. For instance, one could
integrate out the circular shells,
$\Lambda-d\Lambda<{({k_x^2}+{k_y^2})^{1/2}}<\Lambda$, or the
independent rectangular shells,
$\Lambda-d\Lambda<{k_x}<\Lambda$ and $\Lambda-d\Lambda<{k_y}<\Lambda$,
but in either case, one rescales ${k_x}\rightarrow s{k_x}$,
${k_y}\rightarrow s{k_y}$.  It is often convenient to use a regularization
method
which does not explicitly involve a strict limit on the momenta,
but shifts the weight of momentum integrals away from high momenta
in some other fashion.  Then there will be one or more
regulator parameters
which will play a similar role to $\Lambda$.

Of course the overarching concern in the choice of a renormalization
group
transformation is that it leads to a workable calculational procedure.
If one is interested in computing infrared behavior,
this presumably means that it must lead to an infrared fixed point.
The choices available in the scaling of the unintegrated
directions, in the scaling of the fields, etc. should be exercised
in such a way
that the renormalization group
transformation leads to a fixed point. If this
can be done, the renormalization group allows one to relate difficult
low-energy calculations to easy high-energy calculations.
Suppose, for example, that
we have a theory with a dimensionless coupling constant $g$
and kinematics defined by momenta ${p_i}$ some of which
may have cutoffs ${\Lambda_i}$. Suppose further that
we have defined a renormalization group transformation under which some
of the cutoffs ${\Lambda_i}$ have been lowered to
${s^{z_i}}{\Lambda_i}$ in step 1 and that the coupling constant
of the new effective
Lagrangian is $g(s)$. Then the correlation functions satisfy:
$$G({p_i},{\Lambda_i},g) = G({p_i},{s^{z_i}}{\Lambda_i},g(s))\eqn\cutred$$
We now rescale the momenta by
${p_i}\rightarrow {s^{-z_i}}{p_i}$ (including some of the momenta
with no cutoff) and the fields by the appropriate factors to obtain
$$G({p_i},{\Lambda_i},g) = {s^\delta}\,
G({s^{-z_i}}{p_i},{\Lambda_i},g(s))\eqn\scalelaw$$
where ${s^\delta}$ arises from the field rescaling. If the renormalization
group transformation
has an infrared fixed point, $g(s)\rightarrow {g^*}$ as $s\rightarrow 0$,
then we have:
$$G({p_i},{\Lambda_i},g) \rightarrow {s^\delta}\,
G({s^{-z_i}}{p_i},{\Lambda_i},{g^*})\eqn\fpscalelaw$$
The left-hand-side is difficult to calculate directly
in the cases of interest
when the momenta ${p_i}$ are small. In particular, it
is not analytic in $g$, and perturbation theory fails.
However, if we have defined a useful
renormalization group transformation,
then the right-hand-side will be easier to calculate.
This will be the case if the ${s^{-z_i}}{p_i}$'s are comparable
to the ${\Lambda_i}$'s and if loop integrals
really are insensitive to the lack of cutoffs in those directions
which do not have them,
since then $G({s^{-z_i}}{p_i},{\Lambda_i},{g^*})$
will not diverge as the ${p_i}$'s become small.
(If ${g^*}$ is small, one will have the added advantage
of being able to calculate the right-hand-side in low order perturbation
theory -- this is infrared asymptotic freedom, and when it occurs precise
asymptotic results are easily obtained.)
For example, if there is only one
${p_i}$, we
can take ${s^{z_i}}={p_i\over \Lambda}$:
$$G({p_i},{\Lambda_i},g) = {\Bigl({p\over \Lambda}\Bigr)^{\delta/z_i}}\,
G(\Lambda ,\Lambda,g({p^{1/z_i}}))~.\eqn\fpplaw$$
Now
$G(\Lambda ,\Lambda,g({({p\over \Lambda})^{1/z_i}}))$
is just a constant so long as $p$
is small enough that we can neglect the
difference between $g({p^{1/z_i}})$ and ${g^*}$.

\chapter{Flat Fermi Surface}

Let us now consider, in  light of these remarks,
excitations about a flat Fermi surface. The
energy of an excitation is proportional to the distance to the
surface. The free Lagrangian is:
$${S_0} =
\int\,{d\omega\,{d^2}k \biggl\{\psi^{\dagger}\bigl(i\omega
- \epsilon(k)\bigr)\psi\biggr\}}\eqn\free$$
${k_y}$ is the direction perpendicular to the Fermi
surface, and ${k_x}$ is the direction parallel to the
Fermi surface; $\epsilon(k)={v_F}{k_y}$. Following Shankar [\shankar],
and in analogy with the theory of critical dynamics,
we integrate out shells in ${k_y}$ but let $\omega$ range
from $-\infty$ to $\infty$.
After integrating out the high ${k_y}$
modes of $\psi(\omega,k)$, we can rescale $\omega\rightarrow s\omega$,
${k_y}\rightarrow s{k_y}$, and
$\psi \rightarrow s^{-{3\over 2}}\psi$.
But what about ${k_x}$?  The quadratic term \free\ by itself
does not instruct us how to proceed; in particular, it may
or may not be sensible to integrate out high $k_x$ modes, because
they do not necessarily have large energy (unless $k_y$ is also
large.)
In fact, if we do
integrate out these modes, we might be losing track of some
low-energy processes that may be important for
the calculation of certain properties. There is no symmetry
which dictates the scaling of ${k_x}$,
unlike the case of relativistic field theory.  As
long as one is interested in processes that take place in
the neighborhood of a single point on the Fermi surface,
one can integrate out high ${k_x}$ modes and
scale ${k_x}\rightarrow {s^\gamma}{k_x}$ for any $\gamma$;
if one wants to consider processes that
involve distant points, one must take $\gamma=0$.\foot
{Of course it possible for scattering to distant
points on the Fermi surface to occur as a virtual
process; if these are relevant, a full renormalization-group
analysis must include them explicitly.}
For our later purposes it will be useful to consider
non-zero $\gamma$; then, to maintain the form of
the action, we must scale $\psi \rightarrow \psi^{-{3+\gamma}\over 2}$.

Let us consider the scaling of four-Fermi interactions
under this transformation. The term
$${S_4} =
\int\,{{d\omega_1}{d\omega_2}{d\omega_3}\,{d^2}{k_1}{d^2}{k_2}{d^2}{k_3}
\, u({k_{1x}},{k_{2x}},{k_{3x}})\,
{\psi^\dagger}({k_4},{\omega_4})
{\psi^\dagger}({k_3},{\omega_3})
\psi({k_2},{\omega_2})
\psi({k_1},{\omega_1})}\eqn\fourfermi$$
scales as ${s^\gamma}$ (${k_4}={k_1}+{k_2}-{k_3}$ and similarly
for ${\omega_4}$). Hence, it is irrelevant unless $\gamma=0$,
for which case it is marginal.
For any $\gamma$, however, the four-Fermi interaction
is marginal in the kinematic configuration ${k_{1x}}={k_{3x}}$
or ${k_{2x}}={k_{3x}}$.  Thus the Landau parameters
$u({k_{1x}},{k_{2x}},{k_{1x}})$ and $u({k_{1x}},{k_{2x}},{k_{2x}})$ are
marginal -- since there is one fewer $k_x$ integral the scaling
is reduced by ${s^\gamma}$.

This analysis simply demonstrates that fermions at distinct
points near a flat Fermi surface have marginal interactions. As one
focusses on a single point, integrating out processes which occur
far from the point, the only marginal
interactions among the fermions are those
that either preserve or exchange their ${k_x}$ values.

It is instructive to consider the one-loop $\beta$-functional
for these marginal four-Fermi interactions.  It vanishes.
The reason [\shankar ]  for this is that
in the absence of momentum transfer the
internal momenta must be on the Fermi surface.  In an
explicit mode elimination (``Wilsonian'') formulation such momenta
are
not subject to elimination; with other types of
(``field theoretic'')  regulators one still obtains a
null result because the graphs are perfectly finite as the
cutoff is taken to infinity.
This result will come as no suprise to readers familiar with
the literature of Luttinger liquids (\eg\ [\haldanelutt ], [\wenlutt ]:
for one might as well
consider ${k_x}$ here as an
internal quantum number,  and then the system could be interpreted as
a chain of coupled  chiral Luttinger liquids, which are famous for
their marginal interactions.

However, for our present purposes it is more useful
to interpret ${k_x}$ as a direction in momentum
space, so we can introduce transverse gauge fields.
We shall introduce a Chern-Simons gauge field whose
mean field is cancelled and whose fluctuations
are controlled by a ${1\over {|k|^x}}$ interaction as
in
[\hlr,\fixedpoint].\foot{This number $x$, of course, is not to
be confused with the direction $x$. We are following here, as we
did in [\fixedpoint ], the
original notation of [\hlr ].  The reader should be warned that,
lamentably, some authors [\millis ] have chosen to use
the same symbol
$x$ to denote our $1-x$.}
One can use the constraint arising from varying the
vector potential $a_0$ to recast the $1/|k|^2$ repulsion
between fermions into the form
$$S_a~=~\int\,{d\omega}\,{d^2}k\,\epsilon_{ij}\epsilon_{mn}
{k_i}{k_m}k^{-x}{a_j}(k,\omega){a_n}(-k,-\omega)~.\eqn\gaugeterm$$
Assuming that this term dictates the scaling of the vector potentials,
we find $a_x \rightarrow s^{-{1\over 2}(4 + (1-x) \gamma )} a_x$,
$a_y \rightarrow s^{-{1\over 2} (1 + (3-x)\gamma )}a_y$.
(Note that here we have assumed $k_y \ll k_x$, as is appropriate
for $\gamma < 1$.)
We then find that the interaction between the gauge field
and the fermions,
$$ g \int\, {{d\omega}\,{d\omega'}\,{d^2}k\,{d^2}q\,
\psi^{\dagger}(k+q,\omega+\omega')\psi(k,\omega)
{a_i}(q,\omega'){\partial\over{\partial{k_i}}}\epsilon(q+2k)}
\eqn\gfintn$$
scales as $s^{-\gamma(1-x)/2}$. Hence this interaction is
relevant, or at least marginal, so long as $x\leq 1$.

To check this, let us consider the structure of a typical one-loop
graph. Figure 1 shows the one-loop vertex correction.
Its value is:
$${(gv_F)^2}\int{{{d\epsilon}\,{d^2}k}\over{(2\pi)^3}}\,
{1\over{i\epsilon - {v_F}{k_y}}}
\,{1\over{i\omega - i\epsilon - {v_F}({p_y}-{k_y})}}\,
{1\over{{k_x^2}{k^{-x}}}}\eqn\fvertcorr$$
This graph is infrared divergent if the range of integration
includes the origin; thus the cutoff has teeth, and
some of the integrations must be restricted
to a narrow band at the cutoff. It is quite clear that the
${k_x}$ integration is the problem. Restricting the integration
in the ${k_y}$ direction alone is not sufficient because
the denominator of the gauge field propagator can still
become small; it is necessary to restrict the ${k_x}$
integration to a narrow band at the cutoff.

Restricting the ${k_x}$ integration to
$\Lambda-\gamma{d\Lambda}<{k_x}<\Lambda$ is sufficent to make this
integral finite. Furthermore, we can take the $\omega$
and ${k_y}$ cutoffs to infinity and integrate these
variables over their full range in
loop integrals.\foot{This is certainly not the unique choice.
One can integrate out shells of any shape, so long as ${k_x}=0$
is excluded. Of course, the final answers will not depend on
this choice. One should keep the analogy with critical dynamics
in mind. In that case, one integrates out high $k$ modes and
scales to the low $k$, $\omega$ limit; in this case, one
integrates out high ${k_x}$ modes and scales to the low
${k_x}$, ${k_y}$, $\omega$ limit.}
Then, \fvertcorr\ can be evaluated (to lowest order in $1-x$):
$${(gv_F)^2}\int{{{d\epsilon}\,{d^2}k}\over{(2\pi)^3}}\,
{1\over{i\epsilon - {v_F}{k_y}}}
\,{1\over{i\omega - i\epsilon - {v_F}({p_y}-{k_y})}}\,
{1\over{{k_x^2}{k^{-x}}}} =
\gamma\,{{g^2}{v_F}\over{(2\pi)^2}}\,{{d\Lambda}\over\Lambda}
\eqn\fverteval$$
The fermion self-energy diagram may be handled similarly:
$${(gv_F)^2}\int{{{d\epsilon}\,{d^2}k}\over{(2\pi)^3}}\,
{1\over{i\omega - i\epsilon - {v_F}({p_y}-{k_y})}}\,
{1\over{{k_x^2}{k^{-x}}}} =
\omega\,\gamma\,{{g^2}{v_F}\over{2\pi^2}}\,
{{d\Lambda}\over\Lambda}
\eqn\fseeval$$
These may be used to derive recursion relations for $g$ and
${v_F}$,
$${d\over{d\ln\Lambda}}(g{v_F}) = -{\gamma\over2}(1-x)(g{v_F}) +
\gamma\,{{{g^2}{v_F}}\over{(2\pi)^2}}\,(g{v_F})\eqn\ccrecrel$$
$${d\over{d\ln\Lambda}}{v_F} = \gamma\,{{{g^2}{v_F}}\over{(2\pi)^2}}
\, {v_F}\eqn\fvelrecrel$$

In [\fixedpoint], equivalent results were obtained by a field
theoretic technique involving a regularization procedure
similar to dimensional regularization. The pole parts
in $(1-x)$ of the integrals in \fverteval\ and \fseeval\
are cancelled by renormalization counterterms (more
details will be given in the next section). This procedure
is more convenient, particularly for the calculations
of the later sections of this paper, so we will adopt it now.
The one-loop $\beta$-functional and the Fermi velocity
anomalous dimensions may be evaluated by this technique and
one finds:
$$\beta(\alpha) = -{1\over 2}(1-x)\alpha + 4{\alpha^2}\eqn\fbcn$$
$${\eta_{v_F}} = 4\gamma\,\alpha\eqn\vfanomdim$$
where $\alpha=\gamma\,{{{g^2}{v_F}}\over{(2\pi)^2}}$.
$\alpha$ is the correct expansion parameter for perturbation theory,
as may be seen from dimensional analysis. Every divergent loop integral
may be reduced to the form $\int{{dk_x}\over{k_x^{2-x}}}$ which is
a dimension $-(1-x)/2$ quantity; the product of such a term with
$\alpha$ is dimensionless.

\chapter {Curved Fermi Surfaces}

When the Fermi surface is circular, one would like
to impose the additional requirement
of rotational invariance. If ${k_x}$, ${k_y}$ are coordinates
about a point on the Fermi surface as in the flat case, then
the distance above the Fermi surface is
${k_y} + {k_x^2}/2{k_F}$ and hence
$$\epsilon(k) = {v_F}({k_y} + {k_x^2}/2{k_F})\eqn\spenergy$$
for ${k_x}$, ${k_y}$ small compared to ${k_F}$.
To preserve rotational invariance (about the center of the
Fermi circle, not about the origin
of the ${k_x}$, ${k_y}$ coordinates), the
two terms in \spenergy\  should scale
the same way. Hence, we must
take $\gamma={1\over 2}$ [\polchgauge,\fixedpoint].

Otherwise, the circular Fermi surface is completely
analogous to the flat one. It is quite
clear that there is nothing particularly
special about circular Fermi surfaces because
the important fermion-gauge field interactions
occur in the neighborhood of a given point.
\foot{This implies, as well, that these results
apply to the less symmetric Fermi surfaces
of real metals.}
The curvature of the Fermi surface determines
the value of $\gamma$.

In particular, the one-loop
calculations are nearly identical for flat and
curved Fermi surfaces. The presentation
of these calculations given in the previous
section and in [\fixedpoint] is fairly telegraphic,
so in the next section we provide a more detailed and self-contained
analysis, specializing to the case of a circular Fermi surface.
We also take this opportunity to correct some mistakes and
misprints in [\fixedpoint].

\chapter{Effective Action: Screening and Scaling}

We are considering the interacting fermion-Chern-Simons gauge field
system with repulsion.
As before, we shall insert the constraint derived
from varying with respect to $a_0$.  In this way we arrive at the
first six terms of the action we intend to work with:
$$\eqalign{S &=
\int\,{d\omega\,{d^2}k \biggl\{\psi^{\dagger}\bigl(iZ\omega
- Z{Z_{v_F}}\epsilon(k)\bigr)\psi\biggr\}}
+ \int\,{d\omega}\,{d^2}k\,\epsilon_{ij}\epsilon_{mn}
{k_i}{k_m}k^{-x}{a_j}(k,\omega){a_n}(-k,-\omega)\cr &\qquad
+ \int\,{d\omega\,{d^2}k\, {a_0}\epsilon_{ij}{k_i}{a_j}} \cr &\qquad
+ {{\mu}^{{1-x}\over{4}}}
g{Z_g} \int\, {{d\omega}\,{d\omega'}\,{d^2}k\,{d^2}q\,
\psi^{\dagger}(k+q,\omega+\omega')\psi(k,\omega)\,
{a_i}(q,\omega'){\partial\over{\partial{k_i}}}\epsilon(q+2k)}\cr &\qquad
+ {1\over 8}\,{{\mu}^{{1-x}\over{2}}}
{{v_F}\over{k_F}}{g^2}{{Z_g^2}\over{Z{Z_{v_F}}}}
\int\, {{d\omega}\,{d\omega'}\,{d\omega''}\,{d^2}k\,{d^2}q\,{{d^2}q'}\,
\psi^{\dagger}(k+q+q',\omega+\omega'+\omega'')}\cr &\qquad
\times\,{\psi(k,\omega)\,
{a_x}(q,\omega')\,{a_x}(q',\omega'')}\cr &\qquad
+ \int\, {{d\omega}\,{d\omega'}\,{d^2}k\,{d^2}q\,
\psi^{\dagger}(k+q,\omega+\omega')\psi(k,\omega)\,
{a_0}(q,\omega')}\cr &\qquad
+ \int\,{d\omega\,{d^2}k\, {a_0}{a_0}}.\cr}
\eqn\effaction$$

The last, additional term requires some explanation.
It is meant to incorporate
the effect of static screening.  It is standard practice
to include such a term or its equivalent,
at least implicitly, both in this context (\eg\ [\hlr] )
and in the more familiar context of electromagnetism.  In the
latter context, this term parametrizes the plasmon mass.
Two obvious questions it raises are: Why doesn't it violate gauge
invariance? and Where does it come from?  Let us address these in
turn:

Of course, as written, a term $a_0a_0$ in the action {\it does\/}
violate gauge invariance.  However, it really arises in the
form
$$
a_0a_0 ~\sim~ f_{oj} \Pi  f_{oj} ~,
\eqn\polar
$$
involving a polarization operator $\Pi$.
The true polarization operator
is a complicated expression,
even at one loop (the Lindhard function), but it reduces to a
constant times
$1/|k|^2$ at small momenta and frequencies.  For our purpose of
analyzing the infrared behavior, it suffices to keep only the
leading term.  Even within this term we can drop the
$\partial_0 a_j \partial_0 a_j$ terms and the cross terms, because
they are subdominant according to the power-counting that will
presently emerge, at least for $x>0$.

The screening term emerges from the vacuum polarization graph
with the fermions circulating in a loop.  Since it is a loop effect,
there is some logical inconsistency in treating it as part of
the effective action, that we shall then use to generate a
perturbation theory (including its own loops ... ).   However we
must
include this term from the outset, because although
it is formally higher order in the loop expansion or gauge coupling it
is the leading term of its type in the infrared.  Since its purpose
is to remove a singularity at small momenta
that really isn't there, \ie\ to give
the longitudinal part of the gauge field a mass, adding this term
helps stabilize the perturbation scheme.  In principle for
consistency
one should, having stabilized the perturbation
scheme, treat the {\it difference\/} between the
original tree-graph polarization and
the assumed one as an interaction, whose effects could be
assessed perturbatively.  These effects
are presumably small, at least
if the standard
treatment of screening in many-body theory, which seems quite
reasonable
on physical grounds, is correct.  Although we are not aware of
a really adequate discussion along these lines, there is of course
a vast literature on the subject from other points of view
(see [\mahan ]) going back to the classic work of Bohm and
Pines.

While the treatment of screening within effective field theory is
an interesting problem that undoubtedly deserves more attention, we
shall not attempt it here.  In the absence of such a
treatment we cannot escape some looseness in the derivation of
\effaction .  It seems to be the
simplest straightforward implementation of
the standard intuition regarding screening:
that the fermi sea in fact screens, or in other
words removes the source at long wavelength (for us, that means
setting the Chern-Simons magnetic field equal to the fermion density);
and that it generates a plasmon mass.  In any case,
from this point on we will
regard \effaction\ as given, and consider the consequences.

Under the scaling,
$${k_x}\rightarrow {s^{1/2}}{k_x}\eqn\xscaling$$
$${k_y}\rightarrow s{k_y}\eqn\yscaling$$
$${\omega}\rightarrow s{\omega}~,\eqn\tscaling$$
the fields and couplings have the following scaling dimensions:
$$[\psi] = -{7\over4}\eqn\fscaling$$
$$[{a_y}] = -\Bigl({{7-x}\over{4}}\Bigr)\eqn\ayscaling$$
$$[{a_x}] = -\Bigl({{9-x}\over{4}}\Bigr)\eqn\axscaling$$
$$[{a_0}] = -{5\over 4}\eqn\aoscaling$$
$$[{v_F}] = 0\eqn\vscaling$$
$$[g] = -\Bigl({{1-x}\over{4}}\Bigr)~.\eqn\gscaling$$
The scaling of the fermions and the gauge fields is determined
by the condition that the first, second, and final terms
in \effaction\ are left invariant. The fourth term is the
Chern-Simons term which scales as $s^{x/4}$. Although it is
irrelevant in a technical sense, it is the leading $P$, $T$
violating term and is, therefore, important for, e.g
calculations of ${\sigma_{xy}}$ in the $\nu=1/2$ quantum
Hall state. The next two terms are the fermion-gauge field
interactions (the latter of these is required by gauge
invariance in a non-relativistic system). They are irrelevant
for $x>1$, marginal for $x=1$, and relevant for $x<1$.
The final term is the fermion-scalar potential interaction
which is irrelevant since it scales as ${s^{1/4}}$.

The renormalization functions, $Z$, ${Z_{v_F}}$, and
${Z_g}$ in \effaction\ relate the bare and low-energy quantities
($\mu$ is an arbitrary energy scale):
$${\psi_0}={Z^{1/2}}{\psi}\eqn\psibare$$
$${{v_F}\,_0}={Z_{v_F}}{v_F}\eqn\vbare$$
$${g_0}={{{\mu}^{{{1-x}\over{4}}}}g}{{Z_g}\over{ZZ_{v_F}}}
{}~.\eqn\gbare$$
These three functions are sufficient
to cancel all of the divergences in correlation functions
of fermions and gauge fields.

\chapter{Renormalization Group Equations and Their Solution}

One may obtain renormalization
group equations for correlation functions in the standard way.
Differentiating the relationship between bare and renormalized
correlation functions,
$${Z^{n/2}}\,{G^{(n)}}(\omega_i,{v_F}{r_i},\alpha,\mu) =
{G_0^{(n)}}(\omega_i,{v_F}{r_i},\alpha,\mu)\eqn\barecorrelfcn$$
($r$ is the distance to the Fermi surface,
$r={k_y} + {k_x^2}/2{k_F}$ and $\mu$ is the energy scale)
with respect to $\mu$, we find:
$$\Bigl(\mu{\partial\over{\partial\mu}} + \beta(\alpha)
{\partial\over{\partial\alpha}}
+ {n\over2}\eta(\alpha) -
{\eta_{v_F}}(\alpha){r_i}{\partial\over{\partial{r_i}}}\Bigr)
{G^{(n)}}(\omega_i,{v_F}{r_i},\alpha,\mu) = 0\eqn\rgequation$$
since the bare correlation functions
are independent of $\mu$, where
$$\beta(\alpha) = \mu{{d\alpha}\over{d\mu}}\eqn\bfcndef$$
$$\eta(\alpha) = \mu{d\over{d\mu}}\ln Z =
\beta(\alpha){\partial\over{\partial\alpha}}\ln Z\eqn\anomdimdef$$
$${\eta_{v_F}}(\alpha) = \mu{d\over{d\mu}}\ln {Z_{v_F}} =
\beta(\alpha){\partial\over{\partial\alpha}}\ln {Z_{v_F}} \eqn\vfanomdimdef$$
The renormalization group equation, \rgequation, has the
solution:
$${G^{(n)}}(\omega_i,{v_F}{r_i},\alpha,\mu/\lambda) =
{\lambda}^{{n\over2}\eta}
{G^{(n)}}(\omega_i,{v_F}(\lambda){r_i},\alpha(\lambda),\mu)\eqn\rgsoln$$
Combined with the simple scaling property of the Green functions
under the scaling, \xscaling\ - \tscaling,
$${G^{(n)}}(\omega_i,{v_F}{r_i},\alpha,\mu/\lambda) =
{\lambda}^{{7\over4}n - {5\over2}(n-1)}
{G^{(n)}}(\lambda\omega_i,{v_F}\lambda{r_i},\alpha,\mu)\eqn\scalrel$$
this yields
$${G^{(n)}}(\omega_i,{v_F}{r_i},\alpha,\mu) =
{\lambda}^{{5\over2}(n-1) - {7\over4}n + {n\over2}\eta}
{G^{(n)}}({\omega_i}/\lambda,{v_F}(\lambda){r_i}/\lambda,
\alpha(\lambda),\mu)\eqn\rgscalrel$$
For the two-point function, we may take $\lambda=\omega$,
and we find, at low $\omega$,
$${G^{(2)}}(\omega,{v_F}{r},\alpha,\mu) =
{\omega}^{-1 + \eta}
{G^{(2)}}(1,{{{v_F}r}\over{\omega^{1+{\eta_{v_F}}}}},
{\alpha^*},\mu)\eqn\rgscalreltp$$
For correlation functions with insertions of a two-fermion operator,
$O$, we have the renormalization group equation analogous
to \rgequation:
$$\Bigl(\mu{\partial\over{\partial\mu}} + \beta(\alpha)
{\partial\over{\partial\alpha}}
+ {n\over2}\eta(\alpha) + {l}\,{\eta_O}(\alpha) -
{\eta_{v_F}}(\alpha){r_i}{\partial\over{\partial{r_i}}}\Bigr)
{G^{(n,l)}} ({\omega_i},{v_F}{r_i};{\omega_j},{v_F}{r_j},\alpha,\mu) = 0
\eqn\corgequation$$
where
$${\eta_O}(\alpha) = \mu{d\over{d\mu}}\ln {{Z_O}/Z} =
\beta(\alpha){\partial\over{\partial\alpha}}\ln {{Z_O}/Z}\eqn\anomdimdef$$
and $Z_O$ is the counterterm which must be introduced
for the renormalization of correlation functions with $O$ insertions.
The scaling relation which follows from this renormalization group equation,
analogous to the scaling relations, \rgscalrel, is:
$$\eqalign{{G^{(n,l)}}&(\omega_i,{v_F}{r_i};{\omega_j},{v_F}{r_j},\alpha,\mu)
=\cr &
{\lambda}^{{5\over2}(n-1+2l) - {7\over4}(n+2l) - l/4
+ {n\over2}\eta + l{\eta_O}}
{G^{(n,l)}}({\omega_i}/\lambda,{v_F}(\lambda){r_i}/\lambda
;{\omega_j}/\lambda,{v_F}(\lambda){r_j}/\lambda,
\alpha(\lambda),\mu)}\eqn\corgscalrel$$

At the Fermi liquid fixed point, ${\alpha^*}=0$,
which is stable for $x>1$ (or, more generically,
when there are no gauge fields present), these renormalization
group equations
have trivial solutions, such as:
$${G^{(2)}}(\omega,{v_F}{r},\alpha=0,\mu) =
{\omega}^{-1}
{G^{(2)}}(1,{{{v_F}r}\over\omega},
0,\mu)\eqn\fltpfcn$$
$${G^{(0,2)}}(\omega,{v_F}{r},\alpha=0,\mu) =
{G^{(0,2)}}(1,{{{v_F}r}\over\omega},
0,\mu)\eqn\fltpdfcn$$

Since $\alpha$ is a relevant coupling for $x<1$, ${\alpha^*}=0$
is no longer an infrared stable fixed point. In [\fixedpoint],
a new fixed point was found in an expansion in $(1-x)$. An
analogy was drawn between static critical phenomena and this system
in which $(1-x)$ plays the role of $\epsilon=4-d$. The regularization
procedure which was used is analogous to dimensional
regularization: the pole parts in $(1-x)$ of divergent integrals
are cancelled by renormalization countertems.

\chapter{Calculations and Ward Identity}

Following [\fixedpoint],
we will calculate the renormalization
functions and the resulting $\beta$-function by this technique.
We will also say a few words about doing the same calculations
with a cutoff regulator. Such a calculation will be more in the
spirit of the calculations of Halperin, Lee, and Read [\hlr]
(and, of course, more in the spirit of the Wilsonian recursion
relations which we discussed in the section on flat Fermi surfaces).
The zero of the $\beta$-function
is then found to leading order in $(1-x)$.
The anomalous dimensions, $\eta$ and ${\eta_{v_F}}$,
which determine the scaling forms of correlation functions of
fermion fields, are calculated to the same order. Later,
we will also use this method to calculate the anomalous dimensions
of composite operators such as the density and current density;
these anomalous dimensions appear in the
scaling forms of correlation functions with these
operators inserted.
All of the one-loop diagrams of this theory are displayed
in Figure 1. The first two diagrams in Figure 1 are
sufficient to determine the renormalization
group functions $Z$, $Z_{v_F}$, and $Z_g$. The divergence in the
third diagram is related to that in the second by gauge invariance;
either one can be used to determine $Z_g$. The fourth
diagram is subleading by a power
of the external frequency and gives a contribution to an
irrelevant operator, not to one of the terms in the action \effaction.
The fifth and sixth diagrams have
has no singular pieces in $(1-x)$.

The first diagram in Figure 1 is the fermion self-energy diagram.
There are contributions coming from the ${a_y}-{a_y}$, ${a_x}-{a_x}$,
${a_0}-{a_0}$, and ${a_0}-{a_i}$  propagators.
Since there is only one transverse gauge
boson in 2+1 dimensions, one may solve for ${a_x}$
in terms of ${a_y}$.  In other words, we should
choose a gauge.  Here we shall calculate in
radiation gauge,
${a_x}= -({q_y}/{q_x}){a_y}$.
This gauge is actually somewhat unnatural from the point
of view of our scaling (although having reached this point,
in calculating graphs we can use any gauge we please).  In
an Appendix we discuss another, more natural class of gauges,
and check explicitly that the anomalous dimensions of interest
do not depend
upon the gauge choice.
In the kinematic
region of interest, ${q_y}\sim {q_x^2}/{k_F}$ -- as enforced
by the pole at this value in the ${q_y}$ integral -- so the
${a_x}-{a_x}$ propagator is suppressed by a factor of
${q_x^2}/{k_F^2}$ compared to the ${a_y}-{a_y}$ propagator.
The contribution from the
${a_0}-{a_0}$ and ${a_0}-{a_i}$ propagators, too, are
suppressed by powers of $q$ as may be seen directly from the
effective Lagrangian, \effaction\ . Hence, we only
need to consider the contribution to the fermion
self-energy coming from the ${a_y}-{a_y}$  propagator.
$${{g^2}{v_F^2}\over{(2\pi)^3}} \int\, {d\epsilon}\,{dq_x}{dq_y}\,
{1\over{q^{2-x}}}\,{1\over{i\omega - i\epsilon - \epsilon(k-q)}}~.
\eqn
\fermse$$
The ${dq_y}$ integral may be done
by contour integration since ${q_y}$ appears linearly in the
denominator of the the fermion propagator. Then $\epsilon$ disappears
from the integrand, and the $d\epsilon$ integral may be done, leaving
$$2\omega\,{\alpha} \int {{dq_x}\over{q_x^{2-x}}}\,=\,
4\omega\,{\alpha}\Bigl({1\over{1-x}}\Bigr) + {\rm finite\, part}
\eqn\divfse$$
where the divergent part of the integral has been evaluated
by taking the pole part in $(1-x)$ in analogy with
dimensional regularization. Since the self-energy contribution
depends only on $\omega$, we may conclude that $ZZ_{v_F} = 1$
and:
$$Z = Z_{v_F}^{-1} =
1 - 4{\alpha}\Bigl({1\over{1-x}}\Bigr) + O({\alpha^2})~.
\eqn\cterm$$

The second diagram in Figure 1 is the vertex correction.
Again, we need only consider the contribution coming
from the exchange of ${a_y}$ gauge bosons,
$${(gv_F)^2}\int{{{d\epsilon}\,{d^2}k}\over{(2\pi)^3}}\,
{1\over{i{\omega_1} + i\epsilon - \epsilon({p_1}+k)}}
\,{1\over{i{\omega_2} - i\epsilon - \epsilon({p_2}-k)}}\,{1\over{k^{2-x}}}
\,=\, {2\alpha} \int {{dk_x}\over{k_x^{2-x}}}~,\eqn
\vertcorr$$
where the ${dk_x}$ and $d\epsilon$ integrals have been done
as in the self-energy integral. Again, the renormalization
counterterm is chosen to cancel the pole part in $(1-x)$,
$${Z_g} = 1 - 4 {\alpha}\Bigl({1\over{1-x}}\Bigr) + O({\alpha^2})
{}~.\eqn\gcterm$$

Differentiating the equation,
$${\alpha_0}={{{\mu}^{{{1-x}\over{2}}}}\alpha}
{{Z_g^2}\over{{Z^2}{Z_{v_F}}}}
\eqn\alphabare$$
with respect to $\ln\mu$, and solving for $\beta(\alpha)=d\alpha/d\ln\mu$,
we have $\beta(\alpha)$ in the convenient form:
$$\beta(\alpha) =  -{1\over 2}(1-x)
\biggl({{\partial}\over{\partial \alpha}}
\ln\Bigl(\alpha{Z_g}^2/({Z^2}{Z_{v_F}})\Bigr)\biggr)^{-1}.
\eqn\bfcnformula$$

Using \cterm\ and \gcterm\ , we find a $\beta$-function
$$\beta(\alpha) = -{1\over 2}(1-x)\,\alpha + 2{\alpha}^2 +
 O({\alpha^3})\eqn\bfcn$$
and anomalous dimensions,
$$\eta_{v_F}(\alpha) = -\eta(\alpha) = \beta(\alpha)
 {{\partial}\over{\partial \alpha}} \ln{Z_{v_F}}
= -2\alpha + O({\alpha^2})~.\eqn\anomdim$$

The physical interpretation of these equations is quite simple.
The effective coupling, $\alpha$, grows at low energies
on dimensional grounds but this growth is cut off by quantum
fluctuations -- {\it i.e}. screening -- so the coupling approaches
a fixed value, ${\alpha^*} = (1-x)/4$.
%%%This has been changed.
Another way to look at this is to observe that, although
there is no divergent vacuum polarization, there is a relative
renormalization between the space and time parts of
the action.  This renormalization, which is an
important possibility for {\it non-relativistic\/}
systems, leads both to the
running of the effective
coupling and the running of the Fermi velocity.
%%%Perhaps one might want to put this sentence in the
%%%section on the Ward identity.
The effective Fermi
velocity falls to zero as the Fermi surface is approached,
$${v_F} \sim {\omega^{|\eta_{v_F}|}}\,{{v_F}\,_0}\eqn\vfatomega$$
and the quasiparticle weight vanishes with the same exponent,
$$Z \sim {\omega^\eta} \sim {\omega^{|\eta_{v_F}|}}.\eqn\qpweight$$
These anomalous dimensions are reflected in the scaling
form of the fermion Green function.
$$G(\omega,r) = G(\omega,{v_F}(\mu)r,\alpha(\mu),\mu) =
{\omega^{-1+\eta}}\,G\biggl(1,{{v_F}r\over{\omega^{1-|{\eta_{v_F}}|}}},
{\alpha^*},1\biggr)\eqn\fgfscalform$$

We can compare the calculations \fermse\ - \gcterm\
with the same calculations
done with a cutoff regulator. Such a formalism will
be useful later when diagrams with exceptional kinematics are
considered. As an example, consider
the self-energy diagram.
$${{g^2}{v_F^2}\over{(2\pi)^3}} \int\, {d\epsilon}\,{dq_x}{dq_y}\,
{1\over{{q^{2-x}}+i\lambda|\epsilon/q|}}
\,{1\over{i\omega - i\epsilon - \epsilon(k-q)}}~.
\eqn
\cutofffermse$$
Here, we have used the one-loop improved inverse gauge field propagator
which includes the effects of Landau damping, as in [\hlr].
(The justification for such a procedure is the same as
that for the inclusion of the screening term in the action)
The $q_y$-integral may be done first as before, to yield:
$${{g^2}{v_F}\over{(2\pi)^2}} \int\, {d\epsilon}\,{dq_x}\,
{1\over{{q^{2-x}}+i\lambda|\epsilon/q|}}\,{\rm sign}(\epsilon - \omega)~.
\eqn
\cutofffermseqy$$
The $\epsilon$-integration then gives:
$$2\alpha {\int_0^{\Lambda_x}}\,{dq_x}\,{1\over{i\lambda}}\,
\ln({q_x^2} + i\lambda\omega)\eqn\cutofffermseeps$$
The remaining $q_x$-integration must be done with a cutoff.
We find an $\omega$-dependent part (to lowest order in $(1-x)$):
$$2\alpha\,\omega\,\ln(i\lambda\omega/{\Lambda_x^2})
\eqn\cutofflncontr$$
This implies a wavefunction renormalization given by:
$$Z = 1 - 4\alpha\,\ln{\Lambda_x}\eqn\cutoffct$$
Other diagrams are similar.

{\it Ward Identity}: An important identity relating
$Z$ and ${Z_g}$ follows from gauge invariance. The Ward
identity following from the conservation of the current is
(in real time and position space for convenience)
$${\partial_\mu}\langle T({j_\mu}(z){\psi^\dagger}(x){\psi}(y))\rangle
= \delta(z-x)\langle T({\psi^\dagger}(x){\psi}(y))\rangle
+ \delta(z-y)\langle T({\psi^\dagger}(x){\psi}(y))\rangle\eqn\ward$$
This equation relates divergences in the vertex function to
those in the fermion two-point function. The divergent
contributions to the left-hand-side are cancelled by ${Z_g}$ while
the divergent contributions to the right-hand-side are
cancelled by $Z$. Hence, the equality \ward\ implies that
$$Z = Z_g\eqn\wardrgfunctions$$

Actually, a little more care is required in a non-relativistic
theory. The term
${\partial_0}\langle T({\rho}(z){\psi^\dagger}(x){\psi}(y))\rangle$
has divergences cancelled by ${Z_g}$,
while the term
${\partial_i}\langle T({j_i}(z){\psi^\dagger}(x){\psi}(y))\rangle$
has divergences cancelled by ${Z_g}{Z_{v_F}}$
because ${j_i}$ has an explicit factor of $v_F$ in its definition.
Similarly, the fermion propagator on the right-hand-side has an
$\omega$-dependent piece with renormalization function $Z$ and
a term proportional to ${v_F}k$ with renormalization
function $Z{Z_{v_F}}$. Equating the $\omega$- and $k$-dependent
terms separately, we have ${Z_g}=Z$ and ${Z_g}{Z_{v_F}}=Z{Z_{v_F}}$
which both yield \wardrgfunctions.  Our explicit evaluations
respect these identities, as they had better.

\chapter{Renormalization of Composite Operators; Applications}

{\it Landau Parameters}: The Landau parameters, the central
quantities in Landau's Fermi liquid theory, are the marginal
four-Fermi couplings $u({k_{1x}},{k_{2x}},{k_{1x}})$
and $u({k_{1x}},{k_{2x}},{k_{2x}})$, as we mentioned earlier.
(The Landau parameter couplings are not restricted to
the neighborhood of single point where the ${k_x}$, ${k_y}$ coordinates
are valid, but this is the only context in
which we will be considering them.) If there were no gauge interactions,
these couplings would be strictly
marginal, i.e. $\beta(u({k_{1x}},{k_{2x}},{k_{1x}}))
=\beta(u({k_{1x}},{k_{2x}},{k_{1x}}))=0$. However,
interactions with a transverse gauge field could, in principle,
cause them to run.

Consider the simplest case, $u({k_{1x}},{k_{2x}},{k_{1x}})={u_0}$,
and introduce a renormalization counterterm ${Z_u}$ for this coupling.
$${u_0} = u\,{Z_u}/{Z^2}\eqn\lprenfunction$$
${Z_u}$ may be calculated from the diagrams in Figure 2. One sees,
by inspection that $({Z_u}-1) = 2({Z_g}-1)$. But then,
we immediately have ${Z_u}/{Z^2} = 1$ to lowest order. As
a result, the Landau parameters do not scale.

In principle, the fact that they are marginal
means that the interactions parametrized
by the Landau parameters should be included in the effective
action.  However, they do not
contribute
to the renormalization of other couplings, at least to one-loop
order, because of their restricted kinematics.  Indeed because
they merely exchange (or, in three dimensions, rotate) momenta they
cannot link low-momentum to high-momentum modes directly.
At higher orders they would occur, through their influence on
the interactions among virtual high-momentum modes.
Thus they do not affect our calculations, to the order we
performed them.

{\it Cooper Pairing and the $2{k_F}$ Vertex.} The diagram
which determines the one-loop $\beta$-function
of the (marginal) Cooper pairing interaction
(which scatters electrons of momenta $p,-p$ to
momenta $k,-k$) is displayed
in Figure 3. This diagram causes an attractive Cooper pairing
interaction to grow logarithmically, while a repulsive
one is driven logarithmically to zero. In the presence
of a gauge field, the second diagram in Figure 3 also appears at one-loop.
This diagram is unlike the usual Cooper pairing diagram
and unlike the other diagrams which we considered earlier
in that it gives a $(\ln\Lambda)^2$ rather than a simple
$\ln\Lambda$ contribution. For this reason, this diagram must be handled
with a little extra care. We will use a variant of the cutoff regularization
that we discussed earlier.

As a warm-up, let us do the Cooper pairing diagram of Fermi
liquid theory (the first diagram in Figure 3). Here and in the gauge
field case to follow, we will take the simplest case
of an $l=0$ Cooper pairing interaction, $V(k,k')=V$.
We will introduce a frequency cutoff,
but $q_y$ is unrestricted as before; $q_x$ is also unrestricted.
The $q_y$ integration may be done immediately to yield:
$$\eqalign{{V^2} \int\,{d\epsilon}\,{dq_x}{dq_y}\,&
{1\over{i(\omega-\epsilon) - \epsilon(k+q)}}\,
{1\over{i(\omega+\epsilon) - \epsilon(-(k+q))}} \cr & =
{{V^2}\over{v_F}} \int\, {dq_x}\,{{d\epsilon}\over{\epsilon}}
(\theta(\omega+\epsilon)-\theta(\omega-\epsilon))}\eqn\cooper$$
The $q_x$-integral is a harmless angular integral, and the
$\epsilon$-integral gives:
$${{V^2}\over{v_F}}\,\ln(\Lambda/\omega)\eqn\coopercontr$$

With the gauge field, we will use a cutoff regulator and
introduce cutoffs for both
$\epsilon$ and $q_x$. As usual, we do the $q_y$-integral first:
$$\eqalign{{{(g{v_F})^2}V} \int\,{d\epsilon}\,{dq_x}{dq_y}\, &
{1\over{i(\omega-\epsilon) - \epsilon(k+q)}}\,
{1\over{i(\omega+\epsilon) - \epsilon(-(k+q))}}
{1\over{{q^{2-x}}+i\lambda|(\Omega+\epsilon)/q|}}\cr &=
{{\alpha}V} \int\, {dq_x}\,{{d\epsilon}\over{\epsilon}}
{1\over{{q^{2-x}}+i\lambda|(\Omega+\epsilon)/q|}}
(\theta(\omega+\epsilon)-\theta(\omega-\epsilon))}\eqn\gaugecooper$$
The $q_x$-integral may be done up to the cutoff giving:
$${{\alpha}V} {\int_\omega^\Lambda}\,{{d\epsilon}\over{\epsilon}}\,\bigl(
\ln({\Lambda_x^2}+i\lambda(\Omega+\epsilon)) - \ln(i\lambda
(\Omega+\epsilon))\bigr)
\eqn\gaugecooperqy$$
The $\epsilon$-integral then gives
$$\eqalign{{{\alpha}V} {\int_\omega^\Lambda}\,{{d\epsilon}\over{\epsilon}}
&\,\bigl(
\ln({\Lambda_x^2}+i\lambda(\Omega+\epsilon)) -
\ln(i\lambda(\Omega+\epsilon))\bigr)
\cr &= {{\alpha}V}(\Phi(i\Lambda/{\Lambda_x^2},2,1)-
\Phi(i(\omega+\Omega)/{\Lambda_x^2},2,1)
+ {(\ln\Lambda)^2} - {(\ln(\omega+\Omega))^2})}\eqn\coopergaugecontr$$
The presence of a $(\ln\Lambda)^2$ term indicates divergent behavior
in both the ultraviolet and the infrared.
%%%This has been changed.
If we hold the $\Lambda_x$ cutoff constant, and vary
the $\Lambda$ cutoff, then the coupling $V$ is
driven to zero as $e^{-(\ln\Lambda)^2}$.

A very similar effect occurs in the renormalization of the $2{k_F}$
vertex. This vertex is important because of its contribution
to the $2{k_F}$ density-density correlation function and its
role in the calculation of the effects of a quanched random distribution
of impurities. Consider the diagram in Figure 4
where the external fermion lines have low energy and distance
from the Fermi surface but differ
by $2{k_F}$. Integrating ${p_y}$ as above, we find an
integral of precisely the same form as \gaugecooper\ above.
$$\eqalign{{{(g{v_F})^2}} \int\,{d\epsilon}\,{dp_x}{dp_y}\, &
{1\over{i(\omega+\epsilon) - \epsilon(k-p+2{k_F})}}\,
{1\over{i(E+\omega+\epsilon) - \epsilon(k-p)}}
{1\over{{p^{2-x}}+i\lambda|\epsilon/p|}}\cr &=
{{\alpha}} \int\, {dp_x}\,{{d\epsilon}\over{\epsilon+E+2\omega}}
{1\over{{p^{2-x}}+i\lambda|\epsilon/q|}}
({\rm sign}(E+\omega+\epsilon)-{\rm sign}(\omega+\epsilon))}\eqn\gaugetkf$$
As before, this will have the $(\ln \Lambda)^2$ form.

{\it Impurity Scattering.} Scattering by isolated impurities
is also marginal in Fermi liquid theory.
If the impurity is non-magnetic and interacts only through
ordinary potential scattering, the $\beta$-function for the
electron-impurity coupling vanishes
to all orders in the absence of gauge interactions.\foot{If
the impurity is magnetic and interacts through spin-flip scattering,
then the $\beta$-function for the electron-impurity coupling
recieves a non-vanishing contribution from
loop effects. Such a system is the subject of the Kondo model.}
To see this, examine the first diagram in Figure 5 which could,
potentially, renormalize impurity scattering in Fermi liquid
theory. The Landau parameter interactions can only permute
the incoming momenta (or rotate them, in $d=3$),
so the loop momenta are completely
constrained. As a result, the diagrams are non-divergent and
vanish if the internal momenta are restricted to a thin
shell at the cutoff. Hence, these diagrams
do not cause the electron-impurity coupling to run although
they may be numerically important, since they can result
in large corrections. The second diagram in Figure 5 is
unique to the non-Fermi liquid theory with gauge
interactions.

Let
the electron-impurity
coupling is represented by the following term in the effective
Lagrangian (we consider the simplest case of rotationally
invariant scattering),
$$\int\,{{d\omega}\,{d^2}{k_1}{d^2}{k_2}
\, {Z_\lambda}\lambda\,
{\psi^\dagger}({k_2},{\omega})
\psi({k_1},{\omega})}~.\eqn\impcoupling$$
As in the case of the Landau parameters, we see by inspection
that ${Z_\lambda}={Z_g}$, so impurity scattering also
does not run. This is essentially because a single impurity
couples to the local density which is not renormalized
as a result of the Ward identity.
However, a quenched random distribution
of impurities may be thought of
as coupling to a $2{k_F}$ correlation function,
which is renormalized.
This could have observable consequences for localization
in the $\nu=1/2$ quantum Hall system.

{\it Other Composite Operators}: The renormalization
functions $Z$, ${Z_{v_F}}$, and ${Z_g}$ are not sufficient
to cancel the divergences which arise in correlation functions
with composite operator insertions. To calculate,
for instance, the density-density correlation function,
we must introduce the renormalization function ${Z_{\rho}}$.
The anomalous dimension, ${\eta_\rho}$ obtained from this renormalization
function may be substituted into the solution \corgscalrel\ of the
renormalization group equation \corgequation\ to give the scaling form:
$$\langle \rho(q,\omega)\rho(-q,-\omega) \rangle \sim
{\omega^{2{\eta_\rho}}}\,{f_\rho}(1,{{v_F}r\over{\omega^{1-|{\eta_{v_F}}|}}})
\eqn\densitycorr$$
However, the anomalous dimensions of $\rho$ vanish, ${\eta_\rho}=0$
as a result of the Ward identity, \ward, so the scaling from
is the same as that of Fermi liquid theory. It also follows
from the Ward identity that the current recieves
the anomalous dimensions of the Fermi velocity, so
$$\langle j(q,\omega)j(-q,-\omega) \rangle \sim
{\omega^{2{\eta_j}}}\,{f_j}(1,{{v_F}r\over{\omega^{1-|{\eta_{v_F}}|}}})
\sim
{\omega^{-2{\eta_{v_F}}}}\,{f_j}(1,{{v_F}r\over{\omega^{1-|{\eta_{v_F}}|}}})
\eqn\currentcorr$$
By similar arguments, the heat current, $j_Q$, which has an extra
factor of ${v_F}k$ compared to $j$, has anomalous dimensions
${\eta_{j_Q}} = -2{\eta_{v_F}}$.

{\it Logarithmic Corrections at $x=1$}: At $x=1$, the interaction is
marginal and the fixed point coupling is ${\alpha^*}=0$. All
anomalous dimensions vanish, but scaling laws receive logarithmic
corrections. These may be calculated by directly integrating
the one-loop $\beta$-function:
$${{d\alpha}\over{d\ln\mu}} = \beta(\alpha) =2{\alpha^2}\eqn\margbfcn$$
which gives, at low energies:
$$\alpha = {1\over{2\ln\Bigl({{\mu_0}\over{\mu}}\Bigr)}}\eqn\margcoupling$$
This may be substituted into
$${{d\ln{v_F}}\over{d\ln\mu}} = -{\eta_{v_F}} = 2\alpha\eqn\vfflow$$
(obtained by differentiating \vbare) and the resulting equation
may be integrated to give:
$${v_F}(\mu) \sim {{v_{F0}}\over{\ln\Bigl({{\mu_0}\over{\mu}}\Bigr)}}\eqn
\vfscaling$$
Logarithmic corrections to the scaling of composite opertors
may be obtained similarly.

\chapter {Equilibrium and Transport Properties}

The invariance of the effective Lagrangian under our scaling,
together with the one-loop calculation of renormalization
functions, may be used to write the scaling form of the
free energy. The temperature dependence is determined
by finite-size scaling, where the inverse temperature,
$\beta$, is the ``size'' in the time direction.
The equilibrium properties of a metal
described by this fixed point follow immediately
from differentiation of the free energy. The transport
coefficients are given, according to the Kubo formulas,
by correlation functions of density and current operators.
These may be calculated from the renormalization group
equations appropriate to correlation functions with
composite operator insertions. Again, the anomalous dimensions
associated with these are restricted by the Ward identities
resulting from gauge invariance.
The calculation of these quantities using renormalization
group methods is explained in
[\physprop].

{\it Equilibrium properties}: The scaling form for the
free energy density is
$$f \sim {\rm\ const.} +
{{{k_F^{d-1}}\,Q({v_F}{\beta^{\eta_{v_F}}},\dots)}\over
{\beta^2}}\eqn\fescalform$$
Since the free energy density for any theory with a Fermi surface
is proportional to ${k_F^{d-1}}$, the ${\beta^2}$
in the denominator is given by dimensional analysis.
$Q$ is a function of all of the couplings, but these may,
in general, be set equal to their fixed point values
(which is zero for most of them). There is one exception,
however: ${v_F}$ can not be set equal to its fixed point
value, namely zero.
When ${v_F}=0$ and ${k_F}$ is held constant,
the energy of fermionic excitations vanishes, so the free energy
diverges. That is, ${v_F}$ is a {\it dangerous irrelevant
parameter}. As a result, we cannot take
$$f \sim {\rm\ const.} +
{{{k_F^{d-1}}\,Q(0, {u_1^*}, {u_2^*}, \dots)}\over
{\beta^2}}\eqn\wrongscalform$$
but must, rather, take
$$f \sim {\rm\ const.} +
{{k_F^{d-1}}\over
{\beta^2}}\,{A\over {{v_F}{\beta^{\eta_{v_F}}}} } \eqn\rightscalform$$
where $Q({v_F}{\beta^{\eta_{v_F}}},{u_1^*},\ldots)
=A/({v_F}{\beta^{\eta_{v_F}}})$.

The specific heat, ${C_V}=T{{{\partial^2}f}\over {\partial T^2}}$
follows:
$${C_V} \sim T^{1-|{\eta_{v_F}}|}\eqn\specheat$$

To derive the compressibility and magnetic susceptibility, the
scaling form \fescalform\ must be generalized to include the
possibility of a variable chemical potential and magnetic field:
$$f \sim {\rm\ const.} + {1\over{\beta^2}}\,
{{k_F^{d-1}}\,Q({v_F}{\beta^{\eta_{v_F}}},\delta\mu\,{\beta^{1-\eta_\rho}},
H\,{\beta^{1-\eta_\rho}},\ldots)} \eqn\genscalform$$
Then, the compressibility is given by
$$\kappa \sim {{{\partial^2}f}\over {\partial\mu^2}}
\sim T^{2\eta_\rho} \eqn\comp$$
and the (spin) magnetic susceptibility by
$$\chi \sim {{{\partial^2}f}\over {\partial H^2}}
\sim T^{2\eta_\rho} \eqn\susc$$
(For the orbital magnetic susceptibility, replace ${\eta_\rho}$
with ${\eta_j}$). The compressibility and susceptibility may also be
calculated from the $\omega\rightarrow 0$ limit
of the density-density and spin-spin correlation functions.

{\it Transport Properties}: The conductivity is given,
according to the Kubo formula, by:
$$\sigma \sim \lim_{\omega\to0}{{d\over{d\omega}}\langle j(q=0,\omega)
j(-q=0,-\omega)\rangle} \eqn\kfcond$$
Naively applying the scaling formula \currentcorr, we find
$$\langle j(q=0,\omega)
j(-q=0,-\omega)\rangle \sim {T^{-2\eta_{v_F}}} {f_j}(\omega/T,0)
\eqn\ncurrscalt$$
However, as was pointed out in [\physprop], more care is required because
of the presence of the dimensionful parameter, ${k_F}$. The correct
scaling law is:
$$\langle j(q=0,\omega)
j(-q=0,-\omega)\rangle \sim {{k_F}\over{T}}\,
{T^{-2\eta_{v_F}}} {f_j}(\omega/T,0)
\eqn\currscalt$$
unless there is an umklapp process that is relevant, in which case
${k_F}/T$ is replaced by ${k_F}/({g^2}/{k_F})=({k_F}/g)^2$
where $g$ is the reciprocal
lattice vector in question. The conductivity scaling law that
follows from \currscalt\ is:
$$\sigma \sim {1\over{T^{2-2|{\eta_{v_F}}|}}}\eqn\condscallaw$$

The thermal conductivity is given by the following
combination of correlation functions:
$$K = {1\over{T^2}}\Bigl({L^{22}} - {{(L^{12})^2}\over{L^{11}}}\Bigr)
\eqn\kftcond$$
where
$${L^{11}} = T\lim_{\omega\to0}{{d\over{d\omega}}\langle j(q=0,\omega)
j(-q=0,-\omega)\rangle} \eqn\loneone$$
$${L^{12}} = T\lim_{\omega\to0}{{d\over{d\omega}}\langle {j_Q}(q=0,\omega)
j(-q=0,-\omega)\rangle} \eqn\lonetwo$$
$${L^{22}} = T\lim_{\omega\to0}{{d\over{d\omega}}\langle {j_Q}(q=0,\omega)
{j_Q}(-q=0,-\omega)\rangle} \eqn\ltwotwo$$
Hence,
$$K \sim T^{1-2\eta_{v_F}}\sigma\eqn\tconscalt$$

\chapter{Ballistic Propagation and a Direct Measure of Effective
Mass}

In a beautiful experiment Goldman, Su, and Jain [\expts ] studied
the ballistic propagation of quasiparticles near half filling.  Exactly
at half filling the quasiparticles are supposed to travel in straight
lines, even though they are electrically charged and subject to a
large magnetic field, according to the theory of Halperin, Lee, and
Read [\hlr ].  The mutual interactions
among the quasiparticles through the
statistical gauge field is supposed to cancel the applied electromagnetic
background field, as discussed in the first paragraph above.
When the magnetic field is close to but not equal to that which
gives half filling, the quasiparticles will feel the difference as
an effective field, and move in cyclotron orbits.  The effective
field can be much smaller than the real magnetic field, so that the
``internal structure'' of the quasiparticles, which exists
on a scale of the actual
magnetic length, is relatively small, and one can in a
useful approximation
regard them as point particles.
Furthermore they obey Fermi statistics,
and one begins to analyze them by
assuming a Fermi surface as a first approximation, as we have discussed
at length above.
Goldman, Su, and Jain demonstrated that
these ideas are at least roughly valid, by exciting the electron liquid
at one point and finding enhanced response at the position of cyclotron
orbits corresponding to momentum $k_F = 2\sqrt {\pi \rho}$, that is
circles of radius $r = k_F /eB_{\rm eff.}$, where
$B_{\rm eff.} = B - (\rho /\pi e )$ (in units with $\hbar =1$.)

On closer analysis, according to the ideas discussed above,
quasiparticle properties, and not only occupation numbers,
are determined
relative to the nominal Fermi surface.  In particular, the effective
mass diverges, and the the
velocity along a cyclotron orbit
vanishes as its inverse, as the momentum
approaches the nominal Fermi surface.  The technique of
Goldman, Su, and Jain allows one to translate this momentum-space
structure into real space.  By further
measuring the {\it time of flight\/}
along the orbits, as a function of their radius, one could
therefore in principle check the most characteristic feature of
the non-Fermi liquid fixed point, that is the divergence of the
effective mass at the nominal Fermi surface, rather directly.

\chapter{The Distinction Between Quasiparticles and Electrons}

Until now, we have had very little to say about the {\it electrons\/} in
the $\nu=1/2$ quantum Hall effect.
The reason for this is that the description in terms of quasiparticles
is very simple -- essentially that of Fermi liquid theory with
logarithmic corrections -- while the electrons are a complicated
bound state of a quasiparticle and two flux tubes. One might
worry that experimental probes couple directly to electrons
rather than to quasiparticles.  However, for the physical properties
considered in Sections 8 and 9, it is sufficient to consider
the quasiparticles.  For thermodynamic properties this is simply
because the quasiparticles are the actual low-lying excitations, and
for
probes that couple to (fictitious gauge neutral)
fermion bilinears such as the current,
energy, or density the distinction between the quasiparticle fields
and the electron, which differ by a singular gauge transformation,
is unimportant.

However, there are some experiments
for which the single-particle properties of electrons
are important, and in the course of acknowledging their
existence we will make a few brief comments
at this point, that we realize are very far
from exhausting the subject.

The distinction between quasiparticles and electrons
is fundmental when one considers coupling the non-Fermi
liquid to the outside world, as in tunneling.  For the
outside world will not accept quasiparticles, but only
electrons.
The electron spectral weight,
or imaginary part of the retarded electron Green function, is
the relevant quantity for the calculation of the tunneling current.
He, Platzman, and Halperin [\he] have calculated the electron
Green function by assuming that an additional electron added
to a $\nu=1/2$ state may be treated as an infinitely massive
charged particle that undergoes no recoil -- much like the
core hole in the X-ray edge effect -- and then using the well known results
of the X-ray edge problem. They justified this treatment of the electron
by appealing to the effects of the
large magnetic field. They found that
the electron spectral weight
is exponentially suppressed at low frequency, going as
$e^{-{\omega_0}/|\omega|}$. As a result, the tunneling current goes
as $e^{-{V_0}/|V|}$ at small voltage.
Kim and Wen [\instantonspecfun] have found qualitatively similar
results (but not exactly the same; their $\omega_0$ differs
from that of He, {\it et al.} by a factor of 2) using
semiclassical techniques. A classical `instanton' solution
corresponding to the creation of a
quasiparticle together with two flux tubes was found.
The Green function was then found from the Euclidean
action for the instanton-anti-instanton process
corresponding to the creation and annihilation
of an electron. The action for creating an electron
for time $\tau$ goes as $\tau^{1/2}$.\foot{The 2+1-dimensional
spacetime may be thought of as a three-dimensional space
and the instanton-anti-instanton pair may be thought
of as a monopole-anti-monopole pair. The action $\tau^{1/2}$
may be thought of as an energy cost $\sim L^{1/2}$ for creating
a monopole-anti-monopole pair separated by a distance $L$.
Such an energy cost would confine monopoles, but more
weakly than the usual $E\sim L$.} This action cost for creating
an electron alone
implies that electrons are bound to flux tubes -- so
quasiparticles are stable against breaking up into
an electron and two flux tubes -- but only
weakly since the cost is $\sim\tau^{1/2}$.
Running of the coupling will alter these calculations
quantitatively,
if there is a non-trivial fixed point, and perhaps qualitatively
if the fixed point is at zero coupling.  This subject
deserves further
investigation.

One can speculate on the possible relevance
of the flux binding for interpreting an interesting
class of ``effective mass''
measurements (different from, although not unrelated
to, the ones discussed in the previous section.)
We refer to observations of
Shubnikov-de Haas oscillations for
small $\Delta B = B-{B_{1/2}}$, analogous to
more familiar such measurements on ordinary charged
Fermi
liquids at small $B$. If electrons
were infinitely strongly bound to flux tubes,
one would expect to find a constant effective mass
as in the Fermi liquid case.
If electrons were not bound to flux tubes
at all, as in the case of non-interacting charged particles at
half-filling, one would not expect any oscillatory behavior
as a function of $(\Delta B)^{-1}$ and hence no effective mass.
For electrons that are weakly bound to flux tubes, one might
expect oscillations in $(\Delta B)^{-1}$ but an effective mass
with a more interesting asymptotic behavior as
$\Delta B\rightarrow 0$ rather than the constant behavior of
a Fermi liquid near $B=0$.  In recent experiments  [\tsui]
such oscillations have been observed,
and if these experiments are taken at face
value it appear that the  effective mass
diverges as a rather strong
{\it power law\/} in $\Delta B$ .
For several reasons, and especially because the
measurements have not been taken very near $\Delta B=0$
where (according to our considerations) the relatively simple
asymptotic behavior applies, it may be premature to attempt
to fit these experiments.
At weak coupling --
as we have at $x=1$ and small $\Delta B$ --
one would expect a naive
analysis to be correct, and it results in a
logarithmic correction to the effective
mass.  If this differs from observations,
it might mean that the theory is fundamentally flawed,
or (perhaps most plausibly) it might indicate that the
experiments have not yet reached the weak-coupling
regime,
or it might be indicate that nonperturbative
effects, of which instantons
are an example, become quantitatively important even at a
relatively weak coupling.   This subject too deserves
further investigation.

\chapter {Relation to Other Work; and Concluding Remark}

A number of other authors have considered theories of
gauge fields interacting with fermionic excitations
about a Fermi surface.
Inspired by the suggestions of
Anderson and collaborators [\anderson ] that spin-charge separation
occurs in the copper-oxides, several authors [\gaugeguys ]
have considered
theories of fermionic spinons interacting with a gauge
field which serves to eliminate the redundancy in the
spinon-holon description. At the one-loop level they
found non-Fermi liquid behavior, but in the early papers it
was quite unclear how the approximations were being controlled.
Meanwhile, Halperin, Lee, and Read [\hlr ]
considered the $\nu = 1/2$ compressible Hall state of electrons
interacting through an interaction, $V(q) = {1\over {q^x}}$,
and made it quite plausible
that it was described by a
gauge theory similar to that proposed for
the copper-oxide superconductors.
Calculating the one-loop correction to the fermion
propagator, they found non-Fermi liquid behavior
which depended on the exponent $x$; at the physical value,
$x=1$ (Coulomb interaction) they found logarithmic corrections
to Fermi liquid theory.

The first attempt
to justify the results of
low-order perturbation theory from a scaling standpoint
was made by Polchinski [\polchgauge],
who invoked a large-$N$ approximation and assumed
the validity of the analogue of Migdal's theorem (that there
is no significant renormalization of the phonon-electron vertex
function)
in this context.
Here $N$ is
the number of fermion species; of course one is ultimately
interested in small finite values of $N$.
Recently, Altshculer, Ioffe, and Millis [\millis] have expanded
on this large $N$ analysis. Also important in this regard
is the work of Kim, Furasaki, Wen, and Lee [\kim], who showed that
the density-density correlation function is
reliably given in perturbation theory because
it recieves no divergent corrections. Varma [\varma] has shown that
the case $x=0$ is marginal in $d=3$ under a scaling analogous
to that presented here and has considered a $(3-d)$-expansion
that is similar in spirit to the $(1-x)$-expansion
presented here.

In this paper, we have elaborated upon the analysis of [\fixedpoint],
where it was shown that the control parameter $x$, introduced
by Halperin, Lee, and Read [\hlr], could be used to find
a fixed point in a $(1-x)$-expansion analogous to the
$\epsilon$-expansion of critical phenomena. No evidence was found in
[\fixedpoint] for the validity of the analogue
of Migdal's theorem
in the large-N limit.  Indeed, because the fixed point coupling,
${\alpha^*}$, is proportional to $N$, all orders in $\alpha$ in
the $\beta$-function scale as the same power of $N$ near the fixed point.
 From our perspective, the
neglect of higher order corrections is only justified
by the smallness of the parameter $(1-x)$.  Where this
parameter is small, our results essentially agree
with those of
Halperin, {\it et al}. [\hlr], and
Altschuler, {\it et al}. [\millis] (in the sense that
our renormalization-group expressions, expanded in perturbation
theory, agree to the appropriate order with their expressions).
The temperature dependence
of important physical properties of systems
in this universality class is determined
by the anomalous dimensions acquired by the Fermi velocity
and by two-fermion
composite operators.  The anomalous dimensions of these
operators are constrained by the Ward identities resulting from
gauge invariance, and in particular, the density
operator receives no anomalous dimensions.  Thus we
also see no conflict with the main substantive claim
of
Kim, {\it et al.} [\kim].
However, we cannot justify, within our framework, more general
claims about the validity of various resummations of perturbation
theory [\millis ] (or perturbation theory itself [\kim ])
when $1-x$ is not small.  For this regime
one needs more powerful techniques --
possibly those suggested by these authors, or possibly, as
we have suggested, ones more analogous
to those used in extrapolating the $\epsilon$ expansion in critical
phenomena to $\epsilon =1$.

There have also been some authors who have
found radically different behavior from that reported
here or found by the above authors.
Kwon, {\it et al.} [\kwon] and
Altschuler, {\it et al.} [\altschuler]
have used bosonization techniques and found
Green functions equivalent to those of a one-dimensional model
with a four-Fermi coupling that is non-local in time.
Kveschenko and Stamp
[\kveschenko]
have found similar results using an eikonal approximation.
Altschuler, Ioffe, and Millis [\millis]  claim that these results
are appropriate to the $N\rightarrow 0$ limit,
but are not valid for finite
$N$.   From our point of view, it is difficult to see how
a bosonization procedure analogous to the
one appropriate in 1+1 dimensions, which maps the interacting fermion
Lagrangian onto a quadratic boson action,
could be equivalent to our analysis which appears, on the face of it,
to be intrinsically higher-dimensional.

If the ideas discussed in this paper correspond to reality,
Nature has presented us with a truly remarkable condensed matter
system, in which one finds gauge fields, a running
coupling constant, and even a version of asymptotic freedom.
One virtue of this situation is that it allows
one, in principle, to extract precise predictions
with controlled estimates of the errors;
we have attempted here to provide formal tools to
begin this process.

\ack We acknowledge the hospitality of the
Harvard University physics department where this manuscript
was completed.

\endpage

\appendix

\noindent{\it Calculations in Other Gauges}

The calculations of section 7 were done in radiation gauge,
which is particularly simple for calculations, but is not very
natural from the point of view of the scaling \xscaling\  -
\gscaling. We
can, instead, do the calculations in the class of gauges
$$\xi{k_x}{a_y} + {k_y}{a_x} = 0\eqn\natgaugechoice$$
which are natural because both terms scale the same way.
In this gauge, the gauge field propagator has an additional
factor $(1+{\xi^2})^{-1}$ and the vertex has an additional factor
$1+\xi$. As a result, we now find that the $\beta$-function
is:
$$\beta(\alpha) = -{1\over 2}(1-x)\,\alpha + 2
{(1+\xi)\over(1+{\xi^2})}{\alpha}^2 +
 O({\alpha^3})\eqn\gaugebfcn$$
and the anomalous dimensions are:
$$\eta_{v_F}(\alpha) = -\eta(\alpha)
= -2{(1+\xi)\over(1+{\xi^2})}\alpha + O({\alpha^2})~.\eqn\gaugeanomdim$$
but the anomalous dimensions at the fixed point are still
$${\eta_{v_F}}({\alpha^*}) = -{1\over2}(1-x)\eqn\gaugefpad$$

\endpage

\refout

\endpage

\end

\epsfysize=.6\vsize\hskip2cm
\vbox to .6\vsize{\epsffile{diagrams1.eps}}\nextline\hskip-1cm
\endpage
\epsfysize=.8\vsize\hskip2cm
\vbox to .8\vsize{\epsffile{diagrams2.eps}}\nextline\hskip-1cm
\endpage
\epsfysize=.8\vsize\hskip2cm
\vbox to .8\vsize{\epsffile{diagrams3.eps}}\nextline\hskip-1cm
\endpage
\epsfysize=.8\vsize\hskip2cm
\vbox to .8\vsize{\epsffile{diagrams4.eps}}\nextline\hskip-1cm
\endpage
\epsfysize=.8\vsize\hskip2cm
\vbox to .8\vsize{\epsffile{diagrams5.eps}}\nextline\hskip-1cm

\end